\documentclass[aps,amssymb,amsmath,twocolumn,showpacs,superscriptaddress,unsortedaddress]{revtex4}
\usepackage{graphicx}

\begin{document}

\title{Revival of electron coherence in a finite-length quantum wire}

\author{Jaeuk U. Kim}%
\affiliation{Department of Physics, Korea Advanced Institute of
Science and Technology, Daejeon 305-701, Korea}%

\author{W.-R. Lee}%
\affiliation{Department of Physics, Korea Advanced Institute of
Science and Technology, Daejeon 305-701, Korea}%

\author{Hyun-Woo Lee}%
\affiliation{PCTP and Department of Physics,
Pohang University of Science and Technology, Pohang, Kyungbuk 790-784, Korea}%

\author{H.-S. Sim} 
\affiliation{Department of Physics, Korea Advanced Institute of
Science and Technology, Daejeon 305-701, Korea}%

\date{\today}

\begin{abstract}

We study the spatial decay of electron coherence due to
electron-electron interaction in a finite-length disorder-free 
quantum wire. Based on the Luttinger liquid theory, we
demonstrate that the coherence length characterizing the exponential decay of the coherence
can vary from region to region, 
and that the
coherence can even revive after the decay. This counterintuitive behavior,
which is in clear contrast to the conventional exponential decay with single
coherence length,
is due to the 
fractionalization of an electron and the finite-size-induced
recombination of the fractions.
\end{abstract}



\maketitle

{\it Introduction.}---
Quantum coherence of a particle wave is responsible for
various quantum phenomena. 
Conventionally,
the coherence of a particle decays exponentially with
time due to scattering with other particles.
This decay ``law'' was observed experimentally in
electron interferometers~\cite{Hansen01a,Roulleau08a}, where the
interference visibility 
decays as $e^{-L / \ell_\phi}$
with the length $L$ of the
interference path. Here constant
$\ell_{\phi}$ is often called the coherence length,
since the
visibility represents how well the coherence is preserved during the
electron propagation along the path~\cite{Seelig01a}.


Electron-electron interaction is known as a dominant scattering
source that induces the decay of the electron coherence (dephasing)
at low temperature. The interaction generates nontrivial
effects~\cite{Pham00a,Steinberg08a,LeHur05a,Trauzettel,Lebedev05a}.
For instance, when an electron is injected to an infinitely long
one-dimensional wire, the interaction splits it into two fractional
charges~\cite{Pham00a}. The charge fractionalization was
experimentally detected~\cite{Steinberg08a}, and is
responsible~\cite{LeHur05a} for the exponential decay of the
coherence in the infinite wire.

In this Letter, we consider a {\em finite} one-dimensional wire and find surprising deviations from the infinite case in the temperature regime where the thermal energy is comparable to or larger than the
discrete level spacing due to the finite-size effect. The coherence length characterizing the exponential decay of the coherence can vary from region to region, even though the wire is homogeneous, and moreover the coherence can even revive after the decay.
We attribute this counterintuitive behavior to the interaction-induced fractionalization of
electrons \cite{Pham00a,Steinberg08a} and to the separation and
recombination of the fractions in the finite-length wire.
This demonstrates that
electron-electron scattering does not occur in a random
phase-averaging fashion, and clarifies the nature of the coherence of
interacting particles.

\begin{figure}[bt]
\begin{center}
\includegraphics[width=0.35\textwidth]{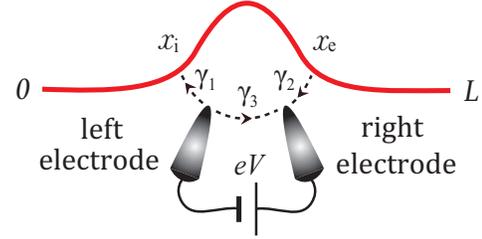}
\caption{(color online) Electron interferometer, consisting of a
disorder-free one-dimensional wire of length $L$ and two electrodes.
Electron tunneling
occurs between the left (right) electrode and the injection position
$x_\textrm{i}$ (extraction $x_\textrm{e}$) of the wire, and between
the electrodes; see dashed arrows.
%
} \label{fig1}
\end{center}
\end{figure}

{\it Interferometer.}--- We consider an electron
interferometer (Fig.~\ref{fig1}), in which a disorder-free wire of length
$L$ weakly couples to two bulk electrodes at two positions
$x_\textrm{i}$ and $x_\textrm{e}$ via electron tunneling.
For simplicity, we ignore the spin degree of freedom for a while, and
neglect the interaction in the electrodes.
The total Hamiltonian of the setup is written as
$H = H_\textrm{wire} + H_L + H_R + H_\textrm{t}$, where
$H_{L(R)}$ is the Hamiltonian of the noninteracting left (right) electrode,
$H_\textrm{t} = [\gamma_1 \Psi^\dag(x_\textrm{i})\Psi_L (0) + \gamma_2 \Psi_R^\dag (0) \Psi(x_\textrm{e})
+ \gamma_3 \Psi_R^\dag (0) \Psi_L (0) + \mathrm{h.c.}]$ describes the tunneling,
$\Psi_{L(R)}(0)$ is the electron field operator at the tunneling point of the left (right) electrode,
and $\gamma_1$, $\gamma_2$, $\gamma_3$ are the tunneling amplitudes along the interference loop.
The electron field operator $\Psi (x)$ at position $x$ in the wire satisfies
$\Psi(0) = \Psi(L)=0$ at the wire boundaries.
The Hamiltonian $H_\textrm{wire}$ of the wire will be given later.

In the setup, under bias voltage $V$, electron current flows between the electrodes via two paths,
the direct tunneling ($\gamma_3$)
and the elastic cotunneling ($\gamma_1 \gamma_2$) through the wire, which cause
the interference.
We derive the interference parts $I_{\rm int}$ of the current,
$I_{\rm int}(x_\textrm{i},x_\textrm{e}) \propto \mathrm{Re}[\gamma_1 \gamma_2
\gamma_3^*] \mathcal{P} \int
d\omega d\omega'
A(x_\textrm{i},x_\textrm{e};\omega)\frac{f_L(\omega')-f_R(\omega')}{\omega'-\omega}$,
by using the Keldysh Green function~\cite{Meir92a,Konig02a} and retaining the
perturbation series up to the lowest order in the tunneling
amplitudes (e.g, for $\gamma_1 \gg \gamma_2, \gamma_3$).
Here
$\mathcal{P}$ means the principal value of the
integral, $f_{L (R)}$ is the Fermi distribution function of the
left (right) electrode, and $A(x_\textrm{i},x_\textrm{e};\omega)$ is a propagator through the wire (introduced below).
The above derivation 
is valid for any specific form of $H_\textrm{wire}$.




In the linear response regime, we obtain
the interference part
$G_\textrm{int} \equiv dI_\textrm{int}/dV$
of the differential conductance,
\begin{eqnarray}
G_\textrm{int} (x_\textrm{i},x_\textrm{e}) \propto
\mathrm{Re}[\gamma_1 \gamma_2 \gamma_3^*] \int_{0}^{\infty} dt
F_T(t) \mathrm{Im}[A(x_\textrm{i},x_\textrm{e};t)].
\label{Conductance}
\end{eqnarray}
Here $A(x_\textrm{i},x_\textrm{e};t) \equiv \langle
\Psi^\dag (x_\textrm{i},0) \Psi (x_\textrm{e},t) + \Psi
(x_\textrm{e},t) \Psi^\dag (x_\textrm{i},0) \rangle_\textrm{w}$ is
the Fourier transform of $A(x_\textrm{i},x_\textrm{e};\omega)$
and represents the electron propagation amplitude in the wire
from $x_\textrm{i}$ to $x_\textrm{e}$ during the time interval $t$,
and
$\langle \cdots \rangle_{\rm w}$ denotes the average over the
equilibrium states of the wire for $\gamma_1=\gamma_2=0$.
The weighting factor, $F_T (t) = \pi k_B T t/[\hbar \sinh (\pi k_B T t/\hbar)]$,
which smears out the interference,
comes from the thermal distribution $f_{L/R}$ of electrons in the
electrodes and from the elastic cotunneling weight $1/(\omega' - \omega)$;
$k_B$ is the Boltzmann constant, $T$ is the temperature,
and $\hbar$ is the Planck constant divided by $2 \pi$. The thermal smearing
is more pronounced for longer $t$, as $F_T(t)$ decays rapidly as
$e^{-\pi k_B T t/\hbar}$ for $t \gg \hbar /(k_B T)$.

All the interaction effects on the coherence are contained in
$A(x_\textrm{i},x_\textrm{e};t)$.
We consider a short-range repulsive interaction.
We evaluate $A$ (thus
$G_\textrm{int}$) by using the bosonization
technique~\cite{Giamarchi04book,Fabrizio95a,Mattsson97a},
a reliable nonperturbative treatment in the low energy regime.
After the bosonization~\cite{Fabrizio95a}, the Hamiltonian of the wire becomes
$H_\textrm{wire} = \epsilon \sum_{q>0} n_q b_q^\dag b_q + \hbar\pi v N^2 / (2gL) $,
where the boson operator $b_q^\dagger$
($[b_q, b_{q'}^\dag]=\delta_{q,q'}$) creates a plasmon with wave vector
$q=\pi n_q/L$ ($n_q = 1,2,\cdots$) and the operator $N$ counts the number of excess electrons in the wire.
Here $\epsilon = \pi \hbar v / L$ is the plasmon level spacing,
$v = v_F / g$ is the plasmon propagation velocity,
$v_F$ is the bare Fermi velocity, and
$g$ is the
Luttinger parameter describing the interaction strength;
$g=1$ in the noninteracting case and $g$ decreases toward $0$ for more repulsive interaction.
The first term of $H_\textrm{wire}$ comes from the
plasmon excitations, while the second from the
zero-mode fluctuations.

\begin{figure}[bt]
\begin{center}
\includegraphics[width=0.44\textwidth]{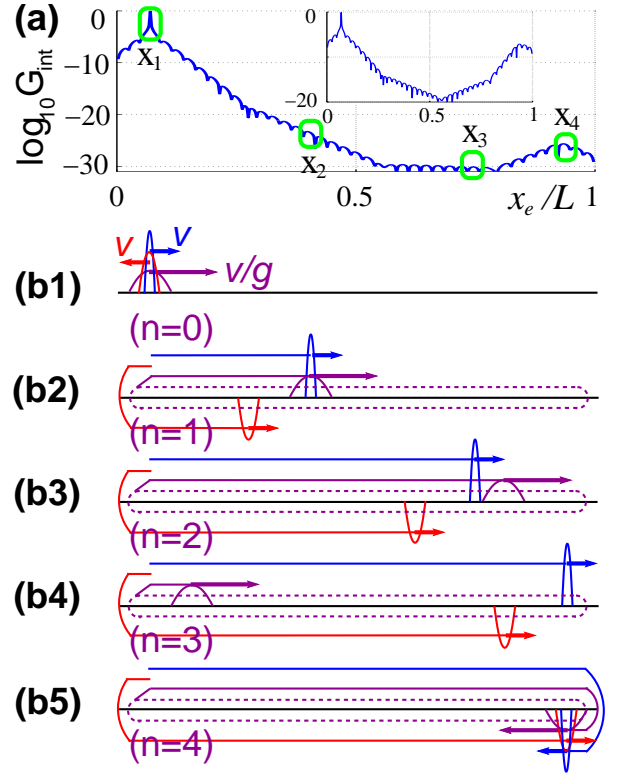}
\caption{{Revival of coherence.} (a) Plot of
$\log_{10} |G_\textrm{int}|$, as a function of 
$x_\textrm{e}$, for the spinless case with $x_\textrm{i} =
0.07L$, $k_B T = 5 \epsilon$, $g = 1/7$, and Fermi wavevector $k_F =
40 \pi / L$. The interference signal
$G_\textrm{int}(x_\textrm{i},x_\textrm{e})$ follows the exponential
decay of $\exp (- |x_\textrm{e}-x_\textrm{i}| / \ell_{\phi})$ with
{\em multiple} coherence lengths $\ell_{\phi}$ as $x_\textrm{e}$
moves from $x_\textrm{i}$, and it {\em revives} around $x_\textrm{e}
= L-x_\textrm{i}$. $G_\textrm{int}$ is normalized by the value at
$x_\textrm{e}=x_\textrm{i}$ and oscillates with period $2 \pi
k_F^{-1}$. Inset: $\log_{10} [|G_\textrm{int}| \exp (|x_\textrm{e} -
x_\textrm{i}| / \ell_{\phi, T})]$. In this plot, the pure thermal
phase-smearing is factored out. (b1 - b5) Schematic views of the
dynamics of the three modes (depicted by blue, red, and purple),
generated at $x_\textrm{i}$ and time $t=0$ by the injection of an
electron to the wire considered in (a). The modes move with
different velocities ($\pm v$ and $v/g$), and are bounced at the
wire boundaries. In (b1-b4), the blue mode arrives at $x_j$ at time
$t_j = (x_j - x_\textrm{i})/v$, $j=1,2,3,4$, moving from
$x_\textrm{i}$ to $x_j$ without any bounce, while in (b5) it arrives
at $x_4$ at $t_5 = (2L - x_4 - x_\textrm{i})/v$ after one bounce at
the right boundary. Here, $x_1 = x_\textrm{i}$, $x_2 = 0.4 L$, $x_3
= 0.75 L$, and $x_4 = L - x_\textrm{i}$ are selected. At each time,
the purple mode has experienced $n$ times of the round trip (the
dashed purple line) with length $2L$. The mode configurations at
time $t_j$ dominantly contribute to
$G_\textrm{int}(x_\textrm{i},x_j)$, and the configurations at $t_4$
and $t_5$ result in the revival of the coherence.} \label{fig2}
\end{center}
\end{figure}

{\it Revival of coherence and multiple coherence lengths}.---
At temperature $k_B T \gtrsim \epsilon$ (the range we focus on),
one might expect that the electron coherence in the finite wire
shows the same exponential decay as in an infinite wire~\cite{LeHur05a},
as the finite level spacing $\epsilon$ 
is masked by $k_B T$.
However our result (Fig.~\ref{fig2}),
obtained from the bosonization and Eq.~\eqref{Conductance}, shows
that finite-size effects persist even in this relatively high temperature regime:
Although $G_\textrm{int}$ follows the
exponential decay form $e^{-|x_\textrm{e} -
x_\textrm{i}|/\ell_\phi}$, the coherence length $\ell_\phi$ changes
from region to region.
Moreover $G_\textrm{int}$ can even have a peak at a special position
$x_\textrm{e}=L-x_\textrm{i}$, showing the revival of the
coherence.


An insight into this striking behavior can be obtained from the bosonization form
of the electron field operator $\Psi (x)$.
For this purpose, we
decompose $\Psi (x)$ into right-moving ($\psi_+$)
and left-moving ($\psi_-$) fields, $\Psi (x) = \psi_+ (x) + \psi_- (x)$,
where $\psi_\pm(x) = (\mp i / \sqrt{2L}) \sum_{k>0} e^{\pm ikx} c_k$.
Here $c_k^\dagger$ creates an electron with wave vector $k=\pi n_k/L$ ($n_k=1,2,\cdots$)
in the wire and satisfies $\{ c_k, c_{k'}^\dagger \} = \delta_{k,k'}$.
The time evolution of $\psi_+$ has the bosonized form~\cite{Fabrizio95a},
\begin{equation}%
\psi_{+}(x,t) \to \frac{e^{i(k_F+\frac{\pi}{2L})x }}{\sqrt{2\pi a}}
e^{i \phi_0(x,t)} e^{i[c_+ \varphi(x- v t) + c_- \varphi(-x-v t)]}.
\nonumber
\end{equation}
Here,
$\phi_0(x,t) = \pi(x-g^{-1}vt)N / L-\chi$ is the fermionic zero mode,
coming from the thermal fluctuation
of the number of electrons occupying the wire,
$c_+ \varphi (x-vt)$ and $c_- \varphi (-x-vt)$ are the bosonic plasmon
modes, $[\chi, N]=i$,
$c_\pm =(g^{-1/2} \pm g^{1/2})/2$, $ \varphi(z) = \sum_{q>0}
\sqrt{\frac{\pi}{qL}} e^{iqz -a q/2} b_{q} + \textrm{h.c.}$, and $a$
is the usual short-distance cutoff.
Note that $\psi_-(x,t)$ has a similar expression.
According to this description, when an electron tunnels into $\psi_+(x_\textrm{i})$,
it
breaks into three fractions in the spinless case, one right-moving
plasmon mode $c_+ \varphi$ (the blue mode of Fig.~\ref{fig2}), another left-moving
plasmon mode $c_- \varphi$ (red), and one right-moving zero mode $\phi_0$
(purple); there is also the tunneling into $\psi_-(x_\textrm{i})$, 
which has the same fractionalization but with ``left'' and ``right''
exchanged. The two plasmon modes move with the same
speed $v$, while the zero mode moves with $v / g$;
in the noninteracting case ($g=1$), no fractionalization occurs, as the blue
and the zero mode move together and the red disappears ($c_-=0$). Because of
bounces at the wire boundaries, the modes separate and recombine
repeatedly.
The
overlap between the modes at time $t$ and the electron state
localized at $x_\textrm{e}$ determines $A(x_\textrm{i},x_\textrm{e};t)$.
The overlap
becomes larger as the
modes locate closer to $x_\textrm{e}$, and drastically enhanced when
some of the modes recombine at $x_\textrm{e}$, 
which is responsible for the nontrivial behavior of the coherence.

We first examine the contribution of the two plasmon modes to the
coherence.
Hereafter we choose $x_\textrm{i} \in [0, L / 2]$ without loss of generality.
The contribution is negligible except for around the times when the
blue mode arrives at $x_\textrm{e}$, since the blue has a bigger
effect on the overlap than the red ($c_+ > c_-$).
In view of the decay in $F_T(t)$ with time $t$, we first consider the shortest one
among those special times, namely $t_\textrm{dire} = |x_\textrm{e}-x_\textrm{i}|/v$
at which the blue mode propagates from $x_\textrm{i}$ to $x_\textrm{e}$
directly without any boundary bounce [Figs.~\ref{fig2}(b1-b4)].
The magnitude of the contribution from $t_\textrm{dire}$ depends on
the separation distance between the blue and red modes at $t_\textrm{dire}$.
A natural candidate of the distance is $d_1 = 2 v t_\textrm{dire}$.
In addition to $d_1$, we have another candidate, $d_2 = 2L - 2 v t_\textrm{dire}$,
which comes from the fact that the two modes recombine at $t= L / v$ ($> t_\textrm{dire}$)
after their boundary bounces [Fig.~\ref{fig2}(b5)].
The smaller of $d_1$ and $d_2$ determines the magnitude of the contribution.
For $x_\textrm{e} \lesssim L/2 + x_\textrm{i}$,
$d_1$ ($< d_2$) increases with $x_\textrm{e}$,
while $d_2$ ($< d_1$) decreases for $x_\textrm{e} \gtrsim L/2 +
x_\textrm{i}$. Thus the contribution from $t_\textrm{dire}$ decreases and then increases
as $x_\textrm{e}$ moves from $x_\textrm{i}$ toward $L$.



The second shortest time is $t_\textrm{boun}= (2L-x_\textrm{i} -
x_\textrm{e} ) / v$ [or $t_\textrm{boun}= (x_\textrm{i} +
x_\textrm{e} ) / v$ if the blue moves to the left], at which time the blue
mode arrives at $x_\textrm{e}$ after one bounce at a wire boundary.
Due to $F_T(t)$, the contribution from $t_\textrm{boun}$ suffers
larger thermal smearing than that from $t_\textrm{dire}$. Unlike
$t_\textrm{dire}$, however, the red mode also arrives at
$x_\textrm{e}$ (recombines with the blue) if
$x_\textrm{e}=L-x_\textrm{i}$, enhancing
$A(x_\textrm{i},x_\textrm{e};t)$ drastically [Fig.~\ref{fig2}(b5)].
Thus around $x_\textrm{e}=L-x_\textrm{i}$, the contributions from
$t_\textrm{dire}$ and $t_\textrm{boun}$ can compete. For smaller $g$
and $x_\textrm{i}$ (closer to wire boundaries), we find that
$t_\textrm{boun}$ becomes more important. In Fig.~\ref{fig2},
$t_\textrm{boun}$ ($t_\textrm{dire}$) is more important for
$x_\textrm{e} \gtrsim L-x_\textrm{i}$ ($x_\textrm{e} \lesssim
L-x_\textrm{i}$). Both the events at $t_{\rm dire}$ and $t_{\rm
boun}$ result in the revival of the coherence around $x_\textrm{e} =
L - x_\textrm{i}$ due to the recombination.


Next we examine the contribution of the zero mode, which is determined by
its overlap with the blue mode arrived at $x_\textrm{e}$.
Since the zero mode moves faster than the blue by factor $1/g$ $(>1)$,
the overlap decays with time right after the electron injection into the wire.
When $x_\textrm{e}$ is sufficiently away from $x_\textrm{i}$, however, it becomes now
possible that the zero mode makes the round trip of the wire once and recombines with
the blue mode [Fig.~\ref{fig2}(b2)], which will suppress the decay.
This recombination and the contribution from the two plasmon modes together result
in the coherence length near $x_\textrm{e}=x_2$ [Fig.~\ref{fig2}(a)], which is different
from the coherence length near $x_\textrm{e} = x_\textrm{i}$.
For sufficiently small $g$, the zero mode experiences the round trip multiple times,
while the blue mode moves directly from $x_\textrm{i}$ to $L - x_\textrm{i}$. Then, the
recombination between the zero mode and the blue can occur at multiple locations. These
multiple recombinations, together with the contributions from the two plasmon modes,
give rise to the multiple coherence lengths in Fig.~\ref{fig2}.

The above interpretation is supported by the following calculation.
We split $A(x_\textrm{i},x_\textrm{e};t)$ into 4 pieces,
$A_{\mu \nu}(x_\textrm{i},x_\textrm{e};t) =
\langle \psi_{\mu}^\dag(x_\textrm{i},0)\psi_{\nu}(x_\textrm{e},t) +
\psi_{\nu}(x_\textrm{e},t) \psi_{\mu}^\dag(x_\textrm{i},0) \rangle_\textrm{w}$,
where $\mu,\nu = +,-$.
At $k_B T \gtrsim \epsilon$ and for $x_\textrm{e} \in [x_\textrm{i},L-x_\textrm{i}]$,
we find that among the pieces, $A_{++}$ dominantly determines $G_\textrm{int}$.
For general $t$, $A_{++} (x_\textrm{i},x_\textrm{e};t)$ is given by
$(\pi a)^{-1} F(x_0) e^{i[k_F + \pi/(2L)] x_-} \textrm{Re} [ e^{i \pi
x_0/(2L)} B(x_\textrm{i},x_\textrm{e};t)]$.
Here, $F(x_0) = \langle e^{i\pi x_0 N/L}\rangle_\textrm{w}$
comes from the zero mode,
%
$ B(x_\textrm{i},x_\textrm{e};t) = [K(x_--vt)]^{c_+^2}
[K(-x_--vt)]^{c_-^2}
 [K(-x_+-vt)K(x_+-vt)]^{c_+ c_-}|K(2x)K(2y)|^{- c_+ c_-}$ is the
plasmon contribution, $x_\pm = x_\textrm{e} \pm x_\textrm{i}$, $x_0
= x_- - g^{-1}vt$, and $ K(z) = (1-e^{-\pi a/L})(1-e^{(iz - a) \pi /
L})^{-1} e^{-4\sum_{q>0} \langle b_{q}^\dag
b_{q}\rangle_\textrm{w} \frac{\pi}{qL}\sin^2\frac{qz}{2}}.$ The
plasmon modes contribute to $A_{++}$ whenever one of the arguments of
$K$'s constituting $B$ vanishes, since $K(z)$ rapidly decreases with
increasing $|z|$ (mod $2L$) at $k_B T \gtrsim \epsilon$.
Among those
times, most important contribution comes from $t_\textrm{dire}$ and
$t_\textrm{boun}$ ($c_+ > c_-$), which are the two shortest
arrival times of the blue mode at $x_\textrm{e}$.





\begin{figure}
\begin{center}
\includegraphics[width=0.42\textwidth]{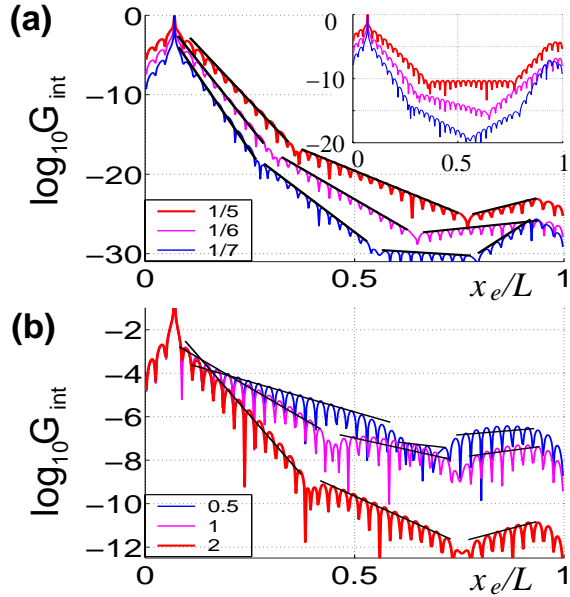}
\caption{{(color online) Dependence on temperature and interaction strength.}
The same as in Fig.~\protect\ref{fig2}(a) except for different
values of $k_B T$ and $g$. (a) $g=1/5$, $1/6$, and $1/7$ from top to
bottom, while $k_B T = 5 \epsilon$ is common. (b) $k_B T = 0.5
\epsilon$, $\epsilon$, and $2 \epsilon$ from top to bottom, while
$g=1/5$. The black lines represent the slopes obtained from $\ell_\phi(n)$
in Eq.~\protect\eqref{Spinless}. }
\label{fig3}
\end{center}
\end{figure}

For $x_\textrm{e}$ around the $n$-th recombination point,
where the zero mode recombines with the blue after the round trip $n$ times,
$A_{++} (x_\textrm{i},x_\textrm{e};t_\textrm{dire})$ is found to be proportional to
\begin{eqnarray}
& & e^{-\ell_{\phi,T}^{-1} [(x_\textrm{e}-x_\textrm{i})
(g+1/g)/2-vt_\textrm{dire} - 2gL[(x_\textrm{e}-x_\textrm{i}-
vt_\textrm{dire}/g)/2L]^2]} \nonumber \\
& \times & e^{-\ell_{\phi,T}^{-1}
[2gL[(x_\textrm{e}-x_\textrm{i}-vt_\textrm{dire}/g)/2L]^2 +2gn
(x_\textrm{e}-x_\textrm{i}- vt_\textrm{dire}/g)]}. \nonumber
\end{eqnarray}
Here, we have used
the approximation of $\langle b_q^\dag b_q \rangle_\textrm{w} \propto 1/q$ in $K(z)$,
which is valid for $k_B T \gtrsim \epsilon$.
The second exponential factor comes from the zero mode while the first exponential factor
describes the overlap between the blue and red plasmon modes.
Note that in the exponents,
the terms quadratic in $x_\textrm{e}-x_\textrm{i}$ cancel with each other while the linear
terms survive. From these linear terms, we find that the coherence length $\ell_\phi(n)$
is given by
\begin{eqnarray}
\ell_\phi^{-1} (n) & = & \ell_{\phi,T}^{-1} + \ell_{\phi,
\textrm{spless}}^{-1} (n), \label{Spinless} \\
\ell_{\phi, \textrm{spless}}^{-1} (n) & = & \ell_{\phi,T}^{-1}
[\frac{g^{-1}+g-2}{2} - 2n(1-g)], \nonumber \\
\ell_{\phi, T} & = & \frac{\hbar v}{\pi k_B T}. \nonumber
\end{eqnarray}
The thermal coherence length $\ell_{\phi, T}$ comes from the thermal
smearing by $F_T$,
while $\ell_{\phi, \textrm{spless}} (n)$ from the interaction
effects;
in the noninteracting case of $g=1$, $\ell_{\phi, T}^{-1}$ still appears,
while $\ell_{\phi, \textrm{spless}}^{-1} (n)$ (thus coherence revival and
multiple coherence lengths) disappears.
The proper values of $n$ and the region where $\ell_\phi
(n)$ is applied depend on $g$ and $x_\textrm{i}$. For $x_\textrm{i}
\lesssim g L / (2-2g)$ [see Fig.~\ref{fig2}], $n$ runs over
$0,1,\cdots,n_\textrm{max}$, where $n_\textrm{max}$ is the largest integer
smaller than $0.5 + (g^{-1}-1) (L-2x_\textrm{i})/(2L)$,
and $\ell_{\phi}(n)$ applies to the
range of $2n-1 \lesssim (g^{-1}-1)(x_\textrm{e}-x_\textrm{i})/L
\lesssim 2n+1$. Equation~\eqref{Spinless} is in excellent agreement with
the calculation of $G_\textrm{int}$ for various values of $g$ and
$T$ (Fig.~\ref{fig3}).
The interaction-induced dephasing
in Eq.~\eqref{Spinless} is caused
by the excitations within energy window of $\sim \pi k_B T$, as the tail of the modes,
which determines the overlap between the modes, decays exponentially
with the rate of $l_{\phi, \textrm{spless}}^{-1} \propto \pi k_B T$.
We remark that the coherence in the finite wire follows the infinite case
only around the injection position, as
the coherence length of the infinite wire~\cite{LeHur05a} is equal to
$\ell_\phi (n=0)$ in Eq.~\eqref{Spinless}, and that
$\ell_\phi (n) \propto T^{-1}$ as in other one dimensional
systems \cite{Hansen01a,Roulleau08a,Seelig01a,LeHur05a}.


From the fact that $\ell_{\phi}(n)$ is negative
in the region where the revival occurs, we find that
the
condition for the occurrence of the revival is
$\ell_{\phi,\rm spless}^{-1}(n_{\rm
max}) <0$. For $x_\textrm{i} \lesssim L/4$, for instance, this
condition results in $g \lesssim 1/3$. When the pure thermal effect
of $\ell_{\phi,T}$ is factored out as in $G_\textrm{int}
e^{|x_\textrm{e}-x_\textrm{i}|/\ell_{\phi,T}}$, the coherence revival
becomes more pronounced (see the insets of
Figs.~\ref{fig2} and \ref{fig3}).


%

%

{\it Spinful case.}---
In the spinful case, the spin mode moves slower than the
charge modes by the factor $g$, showing the spin-charge separation~\cite{Giamarchi04book},
and the interaction is effectively weaker than the spinless case, as
its number of states is two times larger. As a result, for $k_B T
\gtrsim \epsilon$, $\epsilon$ here being the level spacing of the
charge plasmons, we find that Eq.~\eqref{Spinless} is modified into
\begin{eqnarray}
\ell^{-1}_\phi (n) & = & \ell^{-1}_{\phi,T} + \ell^{-1}_{\phi,
\textrm{ch}}(n) +
\ell^{-1}_{\phi, \textrm{sp}}, \label{Spinful} \\
\ell^{-1}_{\phi, \textrm{ch}} (n) & = & \ell^{-1}_{\phi, T} [
\frac{g^{-1} + g -
2}{4} - 2 n (1-g)], \nonumber \\
\ell^{-1}_{\phi, \textrm{sp}} & = & \ell^{-1}_{\phi, T}
\frac{g^{-1} - 1}{2}. \nonumber
\end{eqnarray}
Here, $\ell_{\phi, \textrm{ch}}$ comes from the dynamics of the
charge modes and corresponds to $\ell_{\phi, \textrm{spless}}$,
while $\ell_{\phi, \textrm{sp}}$ shows the dephasing by the
spin-charge separation.
For $x_\textrm{i}/L \lesssim g/(1-g)$, $\ell_\phi (n)$ in
Eq.~\eqref{Spinful} is applied to $2n-1 \lesssim
(g^{-1}-1)(x_\textrm{e}-x_\textrm{i})/(2L) \lesssim 2n+1$, where $n$
runs over $0,1,\cdots,n_\textrm{max}$ and $n_\textrm{max} \simeq
(g^{-1}-1) (L-2x_\textrm{i})/(4L)$. The revival of the coherence appears
at $x_\textrm{e} = L - x_\textrm{i}$ for $g \lesssim 1/5$, when
$x_\textrm{i} \lesssim L/4$.
Note that the revival of the coherence with multiple
coherence lengths due to the charge modes
can be singled out by measuring $G_\textrm{int}
e^{|x_\textrm{e}-x_\textrm{i}| (\ell^{-1}_{\phi,T} +
\ell^{-1}_{\phi, \textrm{sp}})}$.


{\it Conclusion.}---
We have shown that the interplay of the interaction and the
finite-size effect, such as the dynamics of the electron fractionalization
(into the plasmon modes and the zero mode) under the boundary bouncing,
can cause nontrivial behavior of electron coherence in
a finite-size system, which is drastically different from the infinite case.
Our finding
may motivate further research activities towards the understanding
of coherence of interacting particles in various systems~\cite{Neder06a,Youn08a,Neder08a,Sukhorukov}.


We thank M. Bockrath and A. Braggio for discussion.
This work was supported by KRF (2006-331-C00118).

\end{document}